\documentclass{mn2e}
\usepackage{graphicx}

\newcommand{\threej}[6]{\left(\begin{array}{lcr}#1 & #2 & #3 \\ #4 & #5 &
      #6
                    \end{array}\right) }

\newcommand{\sixj}[6]{\left\{\begin{array}{lcr}#1 & #2 & #3 \\ #4 & #5 &
      #6
                    \end{array}\right\} }

\title{Electron-impact rotational and hyperfine excitation of HCN,
HNC, DCN and DNC}

\author[A. Faure, H. N. Varambhia, T. Stoecklin and J. Tennyson]
{Alexandre Faure,$^1$ Hemal N. Varambhia,$^2$ Thierry Stoecklin$^3$ and 
Jonathan Tennyson$^2$\thanks{
Author to whom correspondence should be addressed; 
email:afaure@obs.ujf-grenoble.fr}\\
  $^1$Laboratoire d'Astrophysique, UMR 5571 CNRS, Universit\'e
     Joseph-Fourier, B.P. 53,
  38041 Grenoble Cedex 09, France\\
  $^2$Department of Physics and Astronomy, University College London,
  Gower Street, London WC1E 6BT\\
 $^3$ Institut des Sciences Mol\'eculaires, UMR 5255 CNRS,
  351 cours de la Lib\'eration, 33405 Talence Cedex, France\\}

\date{Accepted ---.  Received ---; in original form ---}

\pagerange{\pageref{firstpage}--\pageref{lastpage}}
\pubyear{2007}

\begin{document}

\maketitle

\label{firstpage}

\begin{abstract}

  Rotational excitation of isotopologues of HCN and HNC by thermal
  electron-impact is studied using the molecular {\bf R}-matrix method
  combined with the adiabatic-nuclei-rotation (ANR)
  approximation. Rate coefficients are obtained for electron
  temperatures in the range 5$-$6000~K and for transitions among all
  levels up to $J=8$. Hyperfine rates are also derived using the
  infinite-order-sudden (IOS) scaling method. It is shown that the
  dominant rotational transitions are dipole allowed, that is those
  for which $\Delta J=1$. The hyperfine propensity rule $\Delta
  J=\Delta F$ is found to be stronger than in the case of He$-$HCN
  collisions. For dipole allowed transitions, electron-impact rates
  are shown to exceed those for excitation of HCN by He atoms by 6
  orders of magnitude. As a result, the present rates should be
  included in any detailed population model of isotopologues of HCN
  and HNC in sources where the electron fraction is larger than
  10$^{-6}$, for example in interstellar shocks and comets.
\end{abstract}


\section{INTRODUCTION}

Hydrogen cyanide, HCN, and its isomer HNC are among the most abundant
organic molecules in space, from star-forming regions to circumstellar
envelopes and comets. HCN and HNC also belong to the small class of
molecules detected in high-redshift galaxies, along with CO and
HCO$^+$ (Gu\'elin et al. 2007 and references therein). In addition to
thermal emission from various rotational transitions within different
vibrational states at (sub)millimeter and far-infrared wavelengths
(see, e.g., Cernicharo et al. 1996), a few masering lines have been
detected toward several stars (e.g. Lucas \& Cernicharo
1989). Numerous isotopologues have also been identified, in particular
the deuterated species DCN and DNC (e.g Leurini et al. 2006 and
references therein).

One interesting observation concerning rotational spectra of HCN is
that of the 'hyperfine anomalies' (Walmsley, Churchwell \& Fitzpatrick
1982 and references therein). At high resolution it is possible to
resolve the hyperfine components arising from the nitrogen ($^{14}$N)
nuclear spin for transitions arising from low-lying rotational
levels. The hyperfine lines have been found in a number of cases
(e.g. Izumiura, Ukita \& Tsuji 1995; Park, Kim \& Minh 1999; Ahrens et
al. 2002 and references therein) to be not in thermal equilibrium with
each other. These anomalies have been shown to depend on the degree of
thermal overlap, on the opacity and on the collisional rates
(e.g. Guilloteau \& Baudry 1981; Gonz\'alez-Alfonso \& Cernicharo
1993; Turner 2001).

Already thirty years ago it was suggested that when the electron
fraction exceeds $\sim$10$^{-5}$, electron-impact excitation of polar
molecules may be significant in addition to collisions with the most
abundant neutrals (Dickinson et al. 1977). Such conditions can be
found for example in diffuse interstellar clouds and photon-dominated
regions where $n$(e)$/n$(H) can reach a few 10$^{-4}$. Recently Lovell
et al. (2004) devoted their study to the effect of electrons in the
rotational excitation of cometary HCN. They showed that electron
collisions are the dominant mechanism of HCN excitation in the comets
Hale-Bopp and Hyakutake where the electron fraction,
$n$(e)$/n$(H$_2$O), lies in the range $\sim$~10$^{-5}$$-$1. The
authors stated that accounting for electron collisions may thus
alleviate the need for large HCN-H$_2$O cross sections in models that
neglect effects due to electrons. Similar conclusions were drawn in
the case of water (Faure, Gorfinkiel \& Tennyson 2004a and references
therein).

Recently Jimenez-Serra et al. (2006) used rotational emissions of SiO,
HCO$^+$, HCN and HNC to probe electron densities in C-type shocks. In
particular they ascribed differences in the ambient and precursor
components to electron density enhancements during the first stages of
the C-type shock evolution. The dipolar ion HCO$^+$ was found to be a
sensitive tracer of this effect.

A theoretical determination of rotational cross sections and rate
coefficients for e$-$HCN was carried out by Saha et al. (1981) for the
rotationally inelastic transitions $J=0\rightarrow 1, J=1\rightarrow
2$ and $J=0\rightarrow 2$. Cross sections were computed using the
rotational close-coupling method and rate coefficients were obtained
in the temperature range 5$-$100~K. An additional study includes that
of Jain \& Norcross (1985), where the adiabatic-nuclei-rotation (ANR)
approximation was combined with model potentials. The cross sections
were computed for the same set of transitions as in Saha et
al. (1981). In addition to electron scattering, the rotational
excitation of HCN by He as substitute for H$_2$ has been studied by
Green \& Thaddeus (1974), Green (unpublished data)\footnote{The
earlier calculations of Green \& Thaddeus (1974) (restricted to the
lowest 8 levels of HCN and temperatures of 5$-$100 K) were extended in
1993 to obtain rate constants among the lowest 30 rotational levels
and for temperatures of 100$-$1200 K. These unpublished results are
available at {\tt http://data.giss.nasa.gov/mcrates/\#hcn}} and
Monteiro \& Stutzki (1986) (see also references therein). This latter
work also provided hyperfine rates for $J$=0$-$4 and $T$=10$-$30~K.

In the present study, the molecular {\bf R}-matrix method has been
combined with the adiabatic-nuclei-rotation (ANR) approximation. We
provide e$-$HCN rotational rates for transitions among all levels up
to $J$=8 and for temperatures in the range 5$-$6000~K. Hyperfine rates
are obtained within the infinite order sudden (IOS) scaling method for
$J=0-3$. Comparisons to the previous theoretical studies and to
differential data is also presented. In Section 2, {\bf R}-matrix
calculations are described and the procedure used to obtain rotational
and hyperfine rates is briefly introduced. In Section 3, both cross
sections and rate coefficients are presented and discussed.
Conclusions are given in Section 4.

\section{CALCULATIONS}

\subsection{R-matrix calculation}
Two of us (Varambhia \& Tennyson 2007) have carried out {\bf R}-matrix
calculations on HCN and HNC at the close-coupling level in C$_{2v}$
symmetry. The calculations were performed using the UK molecular
polyatomic {\bf R}-matrix package (Morgan, Tennyson \& Gillan 1998) at
the (fixed) equilibrium geometries of HCN and HNC using an R-matrix
radius 10~a$_0$.

HCN and HNC (hence DCN and DNC) were represented by the 6-31G Gaussian
type orbital (GTO) basis set.  The target wavefunctions were computed
using the complete active space configuration interaction (CASCI)
method. They were subsequently improved using a pseudo natural
orbitals calculation. The five lowest C$_{2v}$ electronically excited
states $^1$A$_1$, $^3$A$_1$, $^3$A$_2$, $^3$B$_1$, $^3$B$_2$ were
employed and all possible single and double excitations to virtual
orbitals were included. In order to incorporate the double excitation
however, it was necessary to freeze eight electrons (the 1s and 2s
electrons of the C and N atoms). For both HCN and HNC the weighting
coefficients for the density matrix averaging procedure were 5.75,
1.5, 1.5, 1.5, 1.5 for $^1$A$_1$, $^3$A$_1$, $^3$A$_2$, $^3$B$_1$,
$^3$B$_2$ respectively.  This target model yielded dipole moments
$-$2.87~D and $-$2.91~D for HCN and HNC respectively, which can be
compared to the experimental values $-$2.985~D (Ebenstein \& Muenter
1984) and $-$3.05~D (Blackman et al. 1976).

The scattering wavefunctions were represented by a close-coupling
expansion including 24 target electronic states, done to keep the
expected shape resonance as low as possible.  Calculations were
performed on the scattering states $^2$A$_1$, $^2$B$_1$, $^2$B$_2$,
$^2$A$_2$. The continuum GTOs are those of Faure et al. (2002) and are
expanded up to $g$ ($l$=4) partial wave. One virtual orbital was
allocated to each symmetry where target orbitals were available to do
so. This scattering model also used the separable treatment of
continuum and virtual orbitals in order to allow for increased
polarizability and hence further reduce the position of any resonance
yielded by the model.  Further discussion on the problem of
convergence of the polarisation interaction in close-coupling methods
can be found in Varambhia \& Tennyson (2007), Gil et al. (1994) and
Gorfinkiel and Tennyson (2004).  The resulting fixed-nuclei (body
fixed) {\bf T}-matrices were used to obtain the pure rotation
excitation cross-section in the collision energy range 0.01$-$6.2~eV
(see Section~3.2). The positions of the shape resonances for these
molecules are inside the scattering energy ranges.

In the fixed-geometry approximation, the DCN and DNC electronic
wavefunctions are identical to those of HCN and HNC, respectively. The
distinction between the isotopologues therefore arises, in the present
treatment, only from the different rotational excitation thresholds
(see below).

\subsection{Rotational cross sections}

We consider here an electron scattering from a polar linear neutral
molecule in the fixed-nuclei (FN) approximation (Lane 1980). In this
approach, the cross sections are expressed as a partial-wave expansion
within the ANR approximation which assumes that the initial and final
target rotational states are degenerate. For low partial waves (here
$l\leq$4), the cross section is computed from the FN {\bf T}-matrices
obtained via the $R-$matrix calculations. In the case of dipole
forbidden transitions (i.e. those with $\Delta J>1$), cross sections
are expected to converge rapidly and can be safely evaluated using FN
{\bf T}-matrices only (see Section~3.1). In the case of dipole allowed
transitions ($\Delta J=1$), however, the partial-wave expansion does
not converge in the FN approximation, owing to the very long-range
nature of the electron-dipole interaction. To circumvent this problem,
the standard procedure is to use the dipolar Born approximation to
obtain the cross section for the high partial-waves not included in
the FN {\bf T}-matrices (Crawford \& Dalgarno 1971). In this case, the
final cross section is calculated as the sum of two contributions and
can be regarded as a 'short-range correction' to the (dipolar) Born
approximation.

The known unphysical behaviour of the FN cross-sections near
rotational thresholds, inherent in the ANR approximation, is corrected
using a simple kinematic ratio (Chang \& Temkin 1970) which forces the
excitation cross sections to zero at threshold. In the case of
e$-$H$_2$, this procedure has been shown to be accurate down to a
collision energy $E\sim 2\times\Delta E$ where $\Delta E$ is the
rotational threshold (Morrison 1988). Recently, experimental data for
the scattering of cold electrons with water has also confirmed the
validity of the adiabatic 'threshold-corrected' approximation down to
very low electron energies ({\v C}ur{\'{\i}}k et al. 2006; Faure,
Gorfinkiel \& Tennyson 2004b). Note that threshold effects could only
be included rigorously in a full rotational close-coupling
calculation, which is impractical at the collision energies
investigated here. In this context, it is worth stressing that in the
case of molecular ions, cross sections are large and finite at
threshold with a significant but moderate contribution from
closed-channels (Faure et al. 2006).

Finally, in order to make comparisons with available experimental
data, differential cross sections were computed using the Born-closure
method for the scattering amplitude, as described in Itikawa
(2000). In this approach, the high-partial waves contribution ($l>4$)
due to the quadrupole and induced-dipole interactions was also
included.

\subsection{Hyperfine rates}

The hyperfine interaction arises due to the very weak coupling of
nuclear spin to the molecular rotation, which to an excellent
approximation does not affect the overall collision dynamics.
In the laboratory, Ahrens et al. (2002) measured ground state
rotational transitions of HCN using sub-Doppler saturation
spectroscopy in the THz region. This technique enables features such
as hyperfine structure to be revealed, which within Doppler limits,
would remain hidden. Nine consecutive rotational transitions with
their associated hyperfine structure have been thus partly
resolved. Additional study includes that of Turner (2001). Here the
transitions $J=1\rightarrow 0$ and $2\rightarrow 1$ were observed
toward dark clouds for the deuterated molecules N$_2$D$^+$, DCN and
DNC, from which molecular constants including the complex nuclear
quadrupole hyperfine splitting were derived.
More recently the nuclear quadrupole hyperfine structure of HNC and
DNC was resolved in the laboratory for the first time using
millimetre-wave absorption spectroscopy (Bechtel, Steeves \& Field
2006). New rest frequencies for the $J=1\rightarrow 0$, $
J=2\rightarrow 1$ and $J=3\rightarrow 2$ rotational transitions of the
ground vibrational state were determined. It was found that the
hyperfine structure of HNC is dominated by the interaction of the
valence shell electrons with the nuclear spin of the nitrogen
atom. The hyperfine structure of DNC, however, was much more
complicated due to the additional coupling of the deuterium nucleus
($I$=1). This coupling gave rise to a septet in the $J=1\rightarrow 0$
transition.

As hyperfine structure is resolved in some of the astronomical
spectra, it is crucial to know the rate coefficients among these
energy levels. In the following, only the spin of the $^{14}$N atom is
considered as the effects of the spins of H and D atoms on the
collisional rates is assumed negligible, although of course they
affect the energy levels. Hyperfine states are labeled by $J$ and $F$,
where $F$ is the total angular momentum obtained by coupling $J$ to
$I$, the nuclear spin ($I=1$ for $^{14}$N). If the hyperfine levels
are assumed to be degenerate, it is possible to simplify considerably
the hyperfine scattering problem (Corey \& McCourt 1983). Within the
ANR or IOS approximation, which both ignore the rotational spacings,
the scattering equations are further simplified and the rates among
rotational and hyperfine levels can be simply calculated in terms of
`fundamental' rotational rates, $k^{\rm IOS}(L\rightarrow 0)$:
\begin{equation}
\begin{array}{lcl}
k^{\rm IOS}(J\rightarrow J')& = & (2J'+1)\sum_L\threej{J}{J'}{L}{0}{0}{0}^2 \\ 
& & \times (2L+1)k^{\rm IOS}(L\rightarrow 0),
\end{array}
\end{equation}
\begin{equation}
\begin{array}{lcl}
k^{\rm IOS}(JF\rightarrow J'F') & = & (2J+1)(2J'+1)(2F'+1) \\
& & \times\sum_L\threej{J}{J'}{L}{0}{0}{0}^2 \\
& & \times \sixj{L}{F}{F'}{I}{J'}{J}^2 \\
& & \times (2L+1)k^{\rm IOS}(L\rightarrow 0),
\end{array}
\end{equation}
where
\begin{eqnarray}
\threej{J}{J'}{L}{0}{0}{0} {\rm \;and\;} \sixj{L}{F}{F'}{I}{J'}{J}\nonumber
\end{eqnarray}
are Wigner-$3j$ and Wigner-$6j$ symbols, respectively. 

However, as the ANR rotational cross sections are corrected for
threshold effects (see above), Eqs.~(1) and (2) are only moderately
accurate for the actual rates. We have therefore implemented the
'scaling' method proposed by Neufeld \& Green (1994) in which the
hyperfine rates are obtained as a scaling of the rotational rates:
\begin{equation}
k(JF\rightarrow J'F') = \frac{k^{\rm IOS}(JF\rightarrow J'F')}{k^{\rm
IOS}(J\rightarrow J')}k(J\rightarrow J'),
\end{equation}
using the actual rates $k(L\rightarrow 0)$ for the IOS fundamental
rates. For quasi-elastic transitions, i.e. $JF\rightarrow JF'$ with
$F\neq F'$, Eq.~(2) is used directly, as recommended by Neufeld \&
Green (1994). Note that fundamental rates for {\it downward}
transitions were employed as these were found to provide better
results in the case of He$-$HCN for which accurate close-coupling
hyperfine rates are available (Monteiro \& Stutzki 1986).

Finally, it should be noted that within the IOS approximation, the
allowed transitions or selection rules are determined by the
Wigner-$6j$ symbol, as expressed in Eq.~(2). Interestingly, the same
selection rules have been obtained by Chu (1976) and Varshalovich \&
Khersonskii (1977) from multipole expansion approaches. Radiative
(dipolar) selection rules are also determined by the above $6j$ symbol
with $L=1$.

\section{RESULTS AND DISCUSSION}

\subsection{Cross Sections}

The comparison with experimental differential cross sections is often
the only reliable way for testing electron-molecule calculations. In
the case of polar targets, integral cross sections deduced from
experiments are indeed strongly dependent on the extrapolation
procedure used to estimate the small-angle scattering which cannot be
detected experimentally (e.g. Faure et al 2004b).

We have computed the elastic (rotationally summed) differential cross
sections for electron scattering from HCN at the electron energy of
5~eV in order to compare our calculations with the experimental
results of Srivastava et al. (1978). As shown in Fig.~1, our
calculations reproduce the experimental data only
qualitatively. Quantitatively, our results are larger by roughly a
factor of 2 at angles $\theta\leq 100^{\rm o}$. It can be noticed that
the dipolar transition 0$\rightarrow$1 dominates over all other
transitions at both forward and backward angles. Sharp dips are also
observed in the 0$\rightarrow$1 transition at about 20, 70 and
120$^{\rm o}$. We note that these dips, which reflect rainbow
scattering (e.g. Korsch \& Ernesti 1992), are not completely
suppressed by the contribution of the other transitions in the total
curve. More experimental study is clearly necessary to interpret the
difference between theory and experiment, as already suggested
sometime ago by Jain \& Norcross (1985).

\begin{center}
\begin{figure}
\rotatebox{-90}{\resizebox{60mm}{80mm}{\includegraphics*{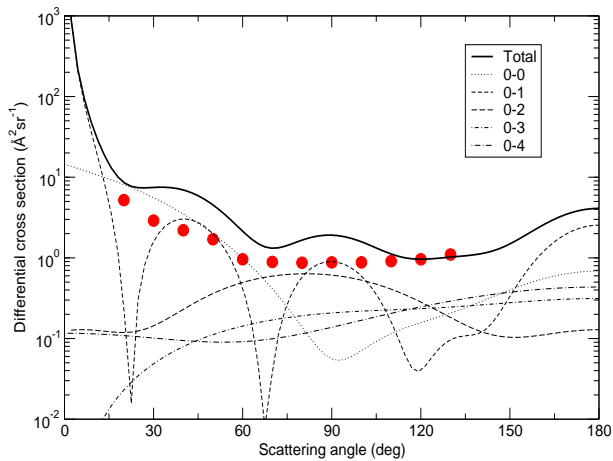}}}
\caption{Differential cross sections for vibrationally elastic
  scattering (rotationally summed) of electrons by HCN at
  5~eV. Experimental data (full circles) are from Srivastava et
  al. (1978). The present calculation is given by the thick solid
  line. Other lines denote partial state-to-state differential cross
  sections.}
\end{figure}
\end{center}

Note that the Born completion was found to be crucial in the case of
the dipole interaction, as expected for very polar species. Dipole
allowed transitions were thus found to be largely dominated by the
high-partial waves ($l>4$) with only a small contribution from the FN
{\bf T}-matrices. In contrast, the Born completion for the quadrupole
and induced-dipole interactions was found to be negligible:
high-partial waves were found to increase cross sections for $\Delta
J=0, 2$ by less than 2\%. This clearly shows that cross sections for
dipole forbidden transitions converge rapidly with respect to the
partial wave expansion, as observed in the case of molecular ions
(e.g. Faure \& Tennyson 2001). As a result, only the Born-closure
approximation for the dipole interaction was implemented in the
computation of the integral rotational cross sections.

In Fig.~2, we compare our integral cross sections with those of Jain
\& Norcross (1985) and Saha et al. (1981). Excellent agreement is
observed down to 0.01~eV. At lower energy, we adopted the
extrapolation formula of Rabad\'an et al. (1998) (Eq.~1 of their
paper) which was calibrated using the rotational close-coupling
results of Saha et al. (1981). This procedure obviously introduces
uncertainties in the rate calculation at temperatures below about
100~K, in addition to closed-channels effects which are completely
ignored in the present treatment. Hence, rate coefficients below 100~K
are expected to be less accurate than at higher temperatures.

\begin{center}
\begin{figure}
\rotatebox{-90}{\resizebox{60mm}{80mm}{\includegraphics*{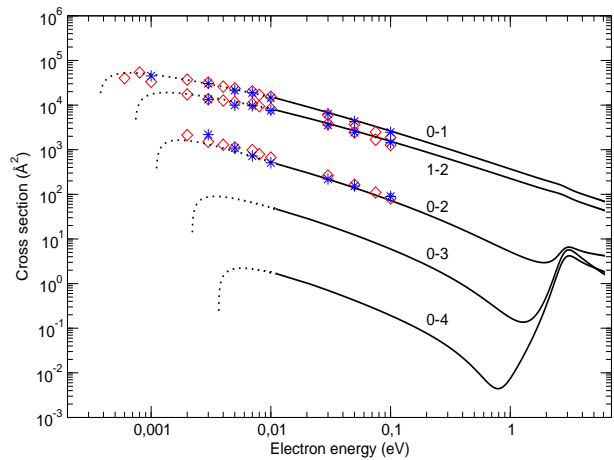}}}
\caption{Integral cross sections for rotationally inelastic scattering
  of electrons by HCN as a function of electron energy. Lozenges and
  stars denotes the results of Saha et al. (1981) and Jain and
  Norcrosss (1985), respectively. The present calculations are given
  by the solid thick lines. Dotted lines are extrapolations (see
  text).}
\end{figure}
\end{center}

A large shape resonance can be observed in Fig.~2 at about 2.8~eV. It
is particularly strong for transitions with small cross sections,
i.e. those with $\Delta J>2$. This resonance (of $^2\Pi$ symmetry) has
also been observed experimentally, as discussed in Varambhia \&
Tennyson (2007). For HNC, it appears at a slightly lower position of
2.5~eV. It is to be noted that the cross sections for $\Delta J=1$ are
much larger than for the other transitions due to the dominance of the
long-range dipole interaction for very polar molecules.

We finally note that, to our knowledge, there is no data (theoretical
or experimental) on the rotational excitation of DCN, HNC and DNC
available in literature.

\subsection{Rotational rates}

As HCN and HNC (hence DCN and DNC) have their first electronic
excitation threshold at 6.63~eV and 6.20~eV, respectively, the
rotational excitation cross sections were computed in the range
0.01$-$6.2~eV and extrapolated down to rotational thresholds (see
Section~3.1). Assuming that the electron velocity distribution is
Maxwellian, rate coefficients were obtained for temperatures in the
range 5$-$6000~K and for transitions among all levels up to
$J$=8. De-excitation rates were calculated using the detailed balance
relation. For later use in modelling, the temperature dependence of
the downward transition rates $k(T)$, in units of cm$^3$s$^{-1}$, was
fitted to the analytic form (Faure et al. 2004a):
\begin{equation}
\log_{10}k(T)=\sum_{i=0}^Na_ix^i
\end{equation}
where $x=1/T^{1/6}$, $N=4$ and $T$ is restricted to the range
5$-$2000~K in order to achieve a fitting accuracy of a few
percent. The coefficients \{a$_i$\}, in units such that $k$ is in
cm$^3$s$^{-1}$, obtained from the fitting procedure are given in
tables 1, 2, 3 and 4 for HCN, DCN, HNC and DNC, respectively. Fitting
coefficients for temperatures above 2000~K and $J>8$ can be obtained
upon request to the authors.


\begin{table*}
\begin{center}
\caption{Coefficients $a_i$ of the polynomial fit, equation~(4), to
  the rotational de-excitation rate coefficients of HCN. These
  coefficients are only valid in the temperature range
  5$-$2000~K. $E_{\rm up}$(K) are the upper level energies.}
\begin{tabular}{crrrrrr}
\hline
Transition  & $E_{\rm up}$(K)  &        $a_0$ &       $a_1$  &         $a_2$&        $a_3$ &        $a_4$ \\         
\hline
$(1-0)$ &      4.3 &      $-$7.543 &       10.018 &     $-$17.974 &       14.328 &      $-$4.169 \\
$(2-0)$ &     12.8 &     $-$10.923 &       22.212 &     $-$52.611 &       60.265 &     $-$26.660 \\
$(2-1)$ &     12.8 &      $-$7.552 &       10.772 &     $-$21.515 &       19.858 &      $-$7.067 \\
$(3-0)$ &     25.5 &     $-$13.297 &       30.136 &     $-$71.602 &       79.988 &     $-$34.340 \\
$(3-1)$ &     25.5 &     $-$10.830 &       22.286 &     $-$52.461 &       59.201 &     $-$25.830 \\
$(3-2)$ &     25.5 &      $-$7.589 &       11.377 &     $-$24.257 &       24.067 &      $-$9.239 \\
$(4-0)$ &     42.5 &     $-$15.647 &       35.452 &     $-$84.761 &       94.380 &     $-$40.338 \\
$(4-1)$ &     42.5 &     $-$13.222 &       30.599 &     $-$73.095 &       81.660 &     $-$35.036 \\
$(4-2)$ &     42.5 &     $-$10.827 &       22.664 &     $-$53.590 &       60.351 &     $-$26.259 \\
$(4-3)$ &     42.5 &      $-$7.633 &       11.910 &     $-$26.550 &       27.495 &     $-$10.976 \\
$(5-0)$ &     63.8 &     $-$17.755 &       36.842 &     $-$88.707 &       99.539 &     $-$42.965 \\
$(5-1)$ &     63.8 &     $-$15.718 &       37.455 &     $-$91.762 &      104.331 &     $-$45.407 \\
$(5-2)$ &     63.8 &     $-$13.277 &       31.655 &     $-$76.702 &       86.569 &     $-$37.456 \\
$(5-3)$ &     63.8 &     $-$10.815 &       22.782 &     $-$53.882 &       60.382 &     $-$26.134 \\
$(5-4)$ &     63.8 &      $-$7.677 &       12.402 &     $-$28.613 &       30.532 &     $-$12.492 \\
$(6-0)$ &     89.3 &     $-$17.947 &       22.002 &     $-$49.290 &       54.613 &     $-$24.371 \\
$(6-1)$ &     89.3 &     $-$18.379 &       44.219 &    $-$114.389 &      136.990 &     $-$62.411 \\
$(6-2)$ &     89.3 &     $-$16.049 &       41.262 &    $-$105.105 &      123.812 &     $-$55.550 \\
$(6-3)$ &     89.3 &     $-$13.426 &       33.410 &     $-$82.796 &       95.273 &     $-$41.916 \\
$(6-4)$ &     89.3 &     $-$10.888 &       23.601 &     $-$56.532 &       63.817 &     $-$27.757 \\
$(6-5)$ &     89.3 &      $-$7.737 &       13.016 &     $-$31.039 &       34.097 &     $-$14.297 \\
$(7-0)$ &    119.1 &     $-$17.477 &        2.037 &        2.640 &     $-$10.257 &        8.971 \\
$(7-1)$ &    119.1 &     $-$19.755 &       39.818 &    $-$106.846 &      130.991 &     $-$59.153 \\
$(7-2)$ &    119.1 &     $-$20.167 &       61.606 &    $-$172.992 &      219.967 &    $-$103.973 \\
$(7-3)$ &    119.1 &     $-$17.003 &       50.704 &    $-$137.567 &      170.812 &     $-$79.770 \\
$(7-4)$ &    119.1 &     $-$13.717 &       36.432 &     $-$93.345 &      110.640 &     $-$49.901 \\
$(7-5)$ &    119.1 &     $-$10.947 &       24.277 &     $-$58.843 &       66.970 &     $-$29.308 \\
$(7-6)$ &    119.1 &      $-$7.791 &       13.571 &     $-$33.249 &       37.364 &     $-$15.963 \\
$(8-0)$ &    153.1 &      $-$9.087 &     $-$41.736 &       42.984 &       14.247 &     $-$28.504 \\
$(8-1)$ &    153.1 &     $-$14.491 &     $-$23.527 &       84.511 &    $-$119.423 &       58.957 \\
$(8-2)$ &    153.1 &     $-$16.613 &       11.279 &     $-$11.836 &      $-$2.489 &        6.619 \\
$(8-3)$ &    153.1 &     $-$17.542 &       36.928 &     $-$88.046 &       95.525 &     $-$39.360 \\
$(8-4)$ &    153.1 &     $-$18.534 &       64.949 &    $-$184.148 &      234.228 &    $-$109.912 \\
$(8-5)$ &    153.1 &     $-$14.334 &       42.448 &    $-$114.006 &      140.539 &     $-$65.328 \\
$(8-6)$ &    153.1 &     $-$11.068 &       25.526 &     $-$63.164 &       73.166 &     $-$32.507 \\
$(8-7)$ &    153.1 &      $-$7.823 &       13.903 &     $-$34.635 &       39.376 &     $-$16.952 \\
\hline
\end{tabular}
\end{center}
\end{table*}

\begin{table*}
\begin{center}
\caption{Coefficients $a_i$ of the polynomial fit, equation~(4), to
  the rotational de-excitation rate coefficients of DCN. These
  coefficients are only valid in the temperature range
  5$-$2000~K. $E_{\rm up}$(K) are the upper level energies.}
\begin{tabular}{crrrrrr}
\hline
Transition  &  $E_{\rm up}$(K)&        $a_0$ &       $a_1$  &         $a_2$&        $a_3$ &        $a_4$ \\         
\hline
$(1-0)$ &      3.5 &      $-$7.513 &        9.737 &     $-$16.754 &       12.437 &      $-$3.174 \\
$(2-0)$ &     10.4 &     $-$10.931 &       22.292 &     $-$52.959 &       60.944 &     $-$27.038 \\
$(2-1)$ &     10.4 &      $-$7.527 &       10.532 &     $-$20.385 &       18.095 &      $-$6.144 \\
$(3-0)$ &     20.9 &     $-$13.275 &       29.938 &     $-$71.022 &       79.434 &     $-$34.150 \\
$(3-1)$ &     20.9 &     $-$10.799 &       22.009 &     $-$51.638 &       58.299 &     $-$25.449 \\
$(3-2)$ &     20.9 &      $-$7.561 &       11.088 &     $-$22.909 &       21.979 &      $-$8.157 \\
$(4-0)$ &     34.8 &     $-$15.559 &       34.612 &     $-$81.872 &       90.381 &     $-$38.336 \\
$(4-1)$ &     34.8 &     $-$13.189 &       30.289 &     $-$72.067 &       80.410 &     $-$34.476 \\
$(4-2)$ &     34.8 &     $-$10.794 &       22.357 &     $-$52.623 &       59.213 &     $-$25.753 \\
$(4-3)$ &     34.8 &      $-$7.592 &       11.502 &     $-$24.778 &       24.819 &      $-$9.611 \\
$(5-0)$ &     52.1 &     $-$17.594 &       35.278 &     $-$83.187 &       91.562 &     $-$38.855 \\
$(5-1)$ &     52.1 &     $-$15.554 &       35.865 &     $-$86.214 &       96.384 &     $-$41.332 \\
$(5-2)$ &     52.1 &     $-$13.199 &       30.907 &     $-$74.137 &       83.071 &     $-$35.733 \\
$(5-3)$ &     52.1 &     $-$10.807 &       22.701 &     $-$53.656 &       60.318 &     $-$26.194 \\
$(5-4)$ &     52.1 &      $-$7.643 &       12.076 &     $-$27.155 &       28.395 &     $-$11.451 \\
$(6-0)$ &     73.0 &     $-$17.202 &       14.785 &     $-$24.098 &       17.750 &      $-$5.212 \\
$(6-1)$ &     73.0 &     $-$17.737 &       38.047 &     $-$92.971 &      105.803 &     $-$46.244 \\
$(6-2)$ &     73.0 &     $-$15.705 &       37.912 &     $-$93.299 &      106.476 &     $-$46.487 \\
$(6-3)$ &     73.0 &     $-$13.236 &       31.583 &     $-$76.490 &       86.276 &     $-$37.307 \\
$(6-4)$ &     73.0 &     $-$10.843 &       23.178 &     $-$55.163 &       62.141 &     $-$27.005 \\
$(6-5)$ &     73.0 &      $-$7.677 &       12.439 &     $-$28.711 &       30.698 &     $-$12.601 \\
$(7-0)$ &     97.3 &     $-$16.289 &      $-$7.875 &       30.337 &     $-$38.294 &       15.702 \\
$(7-1)$ &     97.3 &     $-$18.107 &       24.976 &     $-$59.631 &       69.686 &     $-$32.154 \\
$(7-2)$ &     97.3 &     $-$18.405 &       45.121 &    $-$117.556 &      141.673 &     $-$64.869 \\
$(7-3)$ &     97.3 &     $-$16.011 &       41.257 &    $-$105.090 &      123.775 &     $-$55.517 \\
$(7-4)$ &     97.3 &     $-$13.372 &       33.096 &     $-$81.743 &       93.795 &     $-$41.171 \\
$(7-5)$ &     97.3 &     $-$10.838 &       23.256 &     $-$55.432 &       62.368 &     $-$27.064 \\
$(7-6)$ &     97.3 &      $-$7.728 &       12.953 &     $-$30.725 &       33.628 &     $-$14.069 \\
$(8-0)$ &    125.1 &      $-$9.469 &     $-$41.064 &       53.519 &     $-$24.563 &        5.360 \\
$(8-1)$ &    125.1 &     $-$17.037 &      $-$1.195 &       15.814 &     $-$33.284 &       23.214 \\
$(8-2)$ &    125.1 &     $-$19.699 &       39.962 &    $-$107.384 &      131.888 &     $-$59.701 \\
$(8-3)$ &    125.1 &     $-$19.905 &       59.603 &    $-$166.518 &      211.269 &     $-$99.909 \\
$(8-4)$ &    125.1 &     $-$16.738 &       48.424 &    $-$129.770 &      159.575 &     $-$74.013 \\
$(8-5)$ &    125.1 &     $-$13.600 &       35.446 &     $-$89.911 &      105.637 &     $-$47.303 \\
$(8-6)$ &    125.1 &     $-$10.913 &       24.049 &     $-$58.131 &       66.107 &     $-$28.940 \\
$(8-7)$ &    125.1 &      $-$7.765 &       13.337 &     $-$32.316 &       36.014 &     $-$15.296 \\
\hline
\end{tabular}
\end{center}
\end{table*}


\begin{table*}
\begin{center}
\caption{Coefficients $a_i$ of the polynomial fit, equation~(4), to
  the rotational de-excitation rate coefficients of HNC. These
  coefficients are only valid in the temperature range
  5$-$2000~K. $E_{\rm up}$(K) are the upper level energies.}
\begin{tabular}{crrrrrr}
\hline
Transition &  $E_{\rm up}$(K)&        $a_0$ &       $a_1$  &         $a_2$&        $a_3$ &        $a_4$ \\         
\hline
$(1-0)$ &      4.4 &      $-$7.403 &        8.926 &     $-$14.801 &       10.563 &      $-$2.590 \\
$(2-0)$ &     13.1 &      $-$9.639 &       14.074 &     $-$35.391 &       44.128 &     $-$21.007 \\
$(2-1)$ &     13.1 &      $-$7.423 &        9.788 &     $-$18.744 &       16.659 &      $-$5.761 \\
$(3-0)$ &     26.1 &     $-$12.814 &       28.897 &     $-$74.598 &       89.274 &     $-$40.332 \\
$(3-1)$ &     26.1 &      $-$9.592 &       14.548 &     $-$36.465 &       44.710 &     $-$21.001 \\
$(3-2)$ &     26.1 &      $-$7.455 &       10.360 &     $-$21.423 &       20.817 &      $-$7.926 \\
$(4-0)$ &     43.5 &     $-$15.927 &       39.571 &    $-$100.775 &      117.735 &     $-$52.025 \\
$(4-1)$ &     43.5 &     $-$12.763 &       29.537 &     $-$76.487 &       91.297 &     $-$41.134 \\
$(4-2)$ &     43.5 &      $-$9.578 &       14.777 &     $-$36.902 &       44.619 &     $-$20.689 \\
$(4-3)$ &     43.5 &      $-$7.521 &       11.100 &     $-$24.432 &       25.262 &     $-$10.173 \\
$(5-0)$ &     65.3 &     $-$15.764 &       21.053 &     $-$41.693 &       38.333 &     $-$13.813 \\
$(5-1)$ &     65.3 &     $-$16.129 &       42.754 &    $-$111.610 &      133.075 &     $-$59.844 \\
$(5-2)$ &     65.3 &     $-$12.209 &       24.827 &     $-$61.026 &       71.672 &     $-$32.590 \\
$(5-3)$ &     65.3 &      $-$9.621 &       15.387 &     $-$38.753 &       46.790 &     $-$21.638 \\
$(5-4)$ &     65.3 &      $-$7.583 &       11.785 &     $-$27.233 &       29.445 &     $-$12.313 \\
$(6-0)$ &     91.4 &     $-$15.158 &        1.477 &        7.836 &     $-$15.293 &        7.402 \\
$(6-1)$ &     91.4 &     $-$16.923 &       33.431 &     $-$84.174 &       99.707 &     $-$45.347 \\
$(6-2)$ &     91.4 &     $-$16.830 &       49.989 &    $-$136.387 &      168.824 &     $-$78.263 \\
$(6-3)$ &     91.4 &     $-$13.108 &       33.596 &     $-$90.122 &      110.323 &     $-$50.739 \\
$(6-4)$ &     91.4 &      $-$9.107 &       10.635 &     $-$23.005 &       26.473 &     $-$12.585 \\
$(6-5)$ &     91.4 &      $-$7.668 &       12.643 &     $-$30.602 &       34.339 &     $-$14.756 \\
$(7-0)$ &    121.8 &      $-$9.723 &     $-$58.673 &      181.107 &    $-$236.439 &      111.453 \\
$(7-1)$ &    121.8 &     $-$12.497 &     $-$22.208 &       88.515 &    $-$132.742 &       68.301 \\
$(7-2)$ &    121.8 &     $-$17.644 &       39.253 &     $-$96.958 &      104.695 &     $-$39.340 \\
$(7-3)$ &    121.8 &     $-$18.318 &       64.133 &    $-$183.445 &      234.541 &    $-$110.659 \\
$(7-4)$ &    121.8 &     $-$13.698 &       39.371 &    $-$109.796 &      138.629 &     $-$65.302 \\
$(7-5)$ &    121.8 &      $-$9.814 &       17.433 &     $-$45.443 &       55.767 &     $-$26.026 \\
$(7-6)$ &    121.8 &      $-$7.686 &       12.855 &     $-$31.518 &       35.688 &     $-$15.428 \\
$(8-0)$ &    156.6 &     $-$12.532 &      $-$8.671 &     $-$48.477 &      123.185 &     $-$76.018 \\
$(8-1)$ &    156.6 &      $-$9.784 &     $-$56.529 &      173.568 &    $-$225.735 &      106.060 \\
$(8-2)$ &    156.6 &     $-$12.407 &     $-$22.181 &       87.726 &    $-$131.074 &       67.259 \\
$(8-3)$ &    156.6 &     $-$14.860 &       14.808 &     $-$20.242 &        5.667 &        3.941 \\
$(8-4)$ &    156.6 &     $-$17.161 &       50.763 &    $-$126.624 &      131.935 &     $-$45.525 \\
$(8-5)$ &    156.6 &     $-$14.757 &       49.398 &    $-$143.428 &      186.181 &     $-$89.164 \\
$(8-6)$ &    156.6 &     $-$10.025 &       19.503 &     $-$52.460 &       65.752 &     $-$31.138 \\
$(8-7)$ &    156.6 &      $-$7.546 &        2.952 &        7.712 &     $-$20.936 &       12.334 \\
\hline
\end{tabular}
\end{center}
\end{table*}

\begin{table*}
\begin{center}
\caption{Coefficients $a_i$ of the polynomial fit, equation~(4), to
  the rotational de-excitation rate coefficients of DNC. These
  coefficients are only valid in the temperature range
  5$-$2000~K. $E_{\rm up}$(K) are the upper level energies.}
\begin{tabular}{crrrrrr}
\hline
Transition &  $E_{\rm up}$(K)&        $a_0$ &       $a_1$  &         $a_2$&        $a_3$ &        $a_4$ \\         
\hline
$(1-0)$ &      3.7 &      $-$7.384 &        8.764 &     $-$13.998 &        9.270 &      $-$1.898 \\
$(2-0)$ &     11.0 &      $-$9.621 &       13.929 &     $-$35.013 &       43.803 &     $-$20.891 \\
$(2-1)$ &     11.0 &      $-$7.388 &        9.471 &     $-$17.374 &       14.561 &      $-$4.671 \\
$(3-0)$ &     22.0 &     $-$12.764 &       28.480 &     $-$73.414 &       87.992 &     $-$39.825 \\
$(3-1)$ &     22.0 &      $-$9.554 &       14.223 &     $-$35.527 &       43.683 &     $-$20.576 \\
$(3-2)$ &     22.0 &      $-$7.415 &        9.987 &     $-$19.843 &       18.427 &      $-$6.694 \\
$(4-0)$ &     36.6 &     $-$15.858 &       38.929 &     $-$98.621 &      114.823 &     $-$50.591 \\
$(4-1)$ &     36.6 &     $-$12.724 &       29.182 &     $-$75.374 &       89.958 &     $-$40.533 \\
$(4-2)$ &     36.6 &      $-$9.556 &       14.614 &     $-$36.544 &       44.501 &     $-$20.775 \\
$(4-3)$ &     36.6 &      $-$7.475 &       10.691 &     $-$22.757 &       22.803 &      $-$8.942 \\
$(5-0)$ &     54.9 &     $-$15.543 &       18.882 &     $-$34.000 &       27.004 &      $-$7.873 \\
$(5-1)$ &     54.9 &     $-$15.904 &       40.627 &    $-$104.370 &      122.762 &     $-$54.559 \\
$(5-2)$ &     54.9 &     $-$12.752 &       29.932 &     $-$77.731 &       92.857 &     $-$41.854 \\
$(5-3)$ &     54.9 &      $-$9.558 &       14.818 &     $-$36.983 &       44.601 &     $-$20.644 \\
$(5-4)$ &     54.9 &      $-$7.533 &       11.316 &     $-$25.280 &       26.537 &     $-$10.839 \\
$(6-0)$ &     76.9 &     $-$13.939 &     $-$10.066 &       47.036 &     $-$71.217 &       35.657 \\
$(6-1)$ &     76.9 &     $-$15.950 &       24.200 &     $-$52.614 &       54.255 &     $-$22.087 \\
$(6-2)$ &     76.9 &     $-$16.211 &       44.121 &    $-$116.306 &      139.844 &     $-$63.333 \\
$(6-3)$ &     76.9 &     $-$12.874 &       31.372 &     $-$82.537 &       99.488 &     $-$45.176 \\
$(6-4)$ &     76.9 &      $-$9.596 &       15.308 &     $-$38.466 &       46.324 &     $-$21.378 \\
$(6-5)$ &     76.9 &      $-$7.577 &       11.788 &     $-$27.254 &       29.476 &     $-$12.326 \\
$(7-0)$ &    102.5 &     $-$12.645 &     $-$31.579 &       91.009 &    $-$110.232 &       49.377 \\
$(7-1)$ &    102.5 &     $-$15.491 &        5.896 &      $-$6.381 &        3.094 &      $-$0.486 \\
$(7-2)$ &    102.5 &     $-$17.326 &       37.781 &     $-$98.624 &      119.852 &     $-$55.263 \\
$(7-3)$ &    102.5 &     $-$16.959 &       51.549 &    $-$141.660 &      176.358 &     $-$82.102 \\
$(7-4)$ &    102.5 &     $-$13.109 &       33.807 &     $-$90.846 &      111.375 &     $-$51.288 \\
$(7-5)$ &    102.5 &      $-$9.671 &       16.091 &     $-$40.984 &       49.613 &     $-$22.945 \\
$(7-6)$ &    102.5 &      $-$7.628 &       12.320 &     $-$29.380 &       32.606 &     $-$13.908 \\
$(8-0)$ &    131.8 &     $-$13.106 &      $-$2.647 &     $-$70.842 &      157.263 &     $-$94.056 \\
$(8-1)$ &    131.8 &      $-$9.600 &     $-$58.397 &      180.148 &    $-$235.148 &      110.842 \\
$(8-2)$ &    131.8 &     $-$12.423 &     $-$22.254 &       88.666 &    $-$132.965 &       68.413 \\
$(8-3)$ &    131.8 &     $-$17.403 &       37.141 &     $-$88.825 &       91.193 &     $-$31.327 \\
$(8-4)$ &    131.8 &     $-$18.257 &       63.816 &    $-$182.467 &      233.321 &    $-$110.159 \\
$(8-5)$ &    131.8 &     $-$13.595 &       38.537 &    $-$106.951 &      134.530 &     $-$63.196 \\
$(8-6)$ &    131.8 &      $-$9.779 &       17.182 &     $-$44.598 &       54.590 &     $-$25.434 \\
$(8-7)$ &    131.8 &      $-$7.665 &       12.717 &     $-$31.059 &       35.143 &     $-$15.217 \\
\hline
\end{tabular}
\end{center}
\end{table*}

Excitation rates are presented in Fig.~3 for HCN and DCN. The small
differences with the results of Saha et al. (1981) reflect the
sensitivity of low temperature rates to the near-threshold cross
sections. It can also be noticed that the smaller rotational
thresholds of DCN (with respect to HCN) lead to a significant but
moderate increase of rates below 100~K. As a result, rates for other
isotopologues such as H$^{13}$CN and H$^{15}$NC are expected to be
very similar to those of HCN and HNC, respectively, since the
rotational thresholds are only slightly different.

\begin{center}
\begin{figure}
\rotatebox{-90}{\resizebox{60mm}{80mm}{\includegraphics*{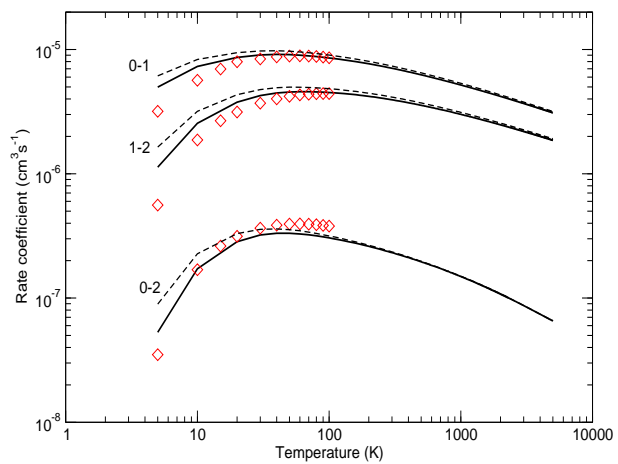}}}
\caption{Rate coefficients for rotational excitation of HCN and DCN by
  electron-impact as a function of temperature. Lozenges denote the
  HCN calculations of Saha et al. (1981) while the solid and dashed
  lines give the present results for HCN and DCN, respectively.}
\end{figure}
\end{center}

In Fig.~4, electron-impact rate coefficients for HCN and HNC are
presented along with rates for excitation of HCN by He atoms. It is
firstly observed that HCN and HNC have similar rates, with typical
differences less than 50~\% for the plotted transitions. Note that the
dipolar $0\rightarrow 1$ curves are almost superposed in the whole
temperature range. This reflects the dominance of the dipole
interaction. Differences larger than a factor of 2 were however found
for the smallest rates, i.e. those corresponding to transitions with
$\Delta J>2$. Second, rate coefficients for excitation by He atoms are
much smaller than electron-impact rates. In particular the He$-$HCN
rate for the transition $0\rightarrow 1$ is about 6 orders of
magnitude smaller than the e$-$HCN rate. Furthermore, the propensity
rules are very different, with even $\Delta J$ strongly favored over
odd $\Delta J$ in the case of He. This propensity reflects an
interference effect related to the even anisotropy of the He$-$HCN PES
(e.g. McCurdy \& Miller 1977). As a result, the difference between
e$-$HCN and He$-$HCN rates is much larger for dipole allowed
transitions than for other transitions. For instance, the He$-$HCN
rate coefficient for the transition $0\rightarrow 4$ at 1000~K is only
a factor of 4 smaller than the electron-impact rate.

These large differences between He$-$HCN and e$-$HCN rotational rates,
both in terms of magnitude and propensity rules, suggest that the
modelling of HCN excitation in astrophysical environments must be very
sensitive to the relative abundance of electrons and neutrals.

\begin{center}
\begin{figure}
\rotatebox{-90}{\resizebox{59mm}{80mm}{\includegraphics*{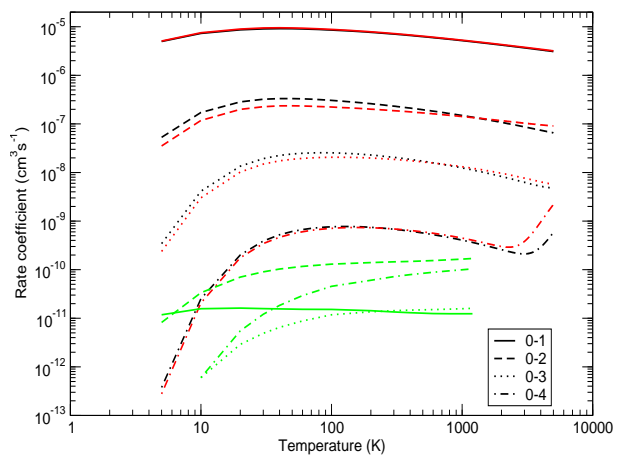}}}
\caption{Rate coefficients for rotational excitation of HCN and HNC as
  a function of temperature. The black and red lines denote the
  present results for electron-impact excitation of HCN and HNC,
  respectively. The green line gives the results of Green \& Thaddeus
  (1974) and Green (unpublished data) for the rotational excitation of
  HCN by He atoms.}
\end{figure}
\end{center}

\subsection{Hyperfine rates}

Hyperfine rate coefficients were computed from the pure rotational
rates using the formalism proposed by Neufeld \& Green (1994), as
explained in Section.~2.3. Sample results are presented in Tables~5,
6, 7 and 8 for HCN, DCN, HNC and DNC, respectively. Additional
hyperfine data at different temperatures and for higher rotational
levels can be obtained upon request to the authors. The relative
propensities predicted by our calculations at 20~K for a few hyperfine
transitions induced by electron-impact are given in Table~9 and are
compared to He$-$HCN results. Firstly, it may be seen that the present
implementation of the method of Neufeld \& Green (1994), i.e. based on
downward fundamental rates, gives a reasonable agreement with the
accurate close-coupling results of Monteiro \& Stutzki (1986) for
He$-$HCN. Errors on rate values were thus found to range between
typically 10 and 50~\%, suggesting that the IOS scaling of rotational
rates provides a rather good description of hyperfine rates. Second,
it is found that the well known propensity rule $\Delta J=\Delta F$ is
stronger for electron$-$HCN than for He$-$HCN. This propensity rule
was derived and confirmed experimentally by Alexander \& Dagdigian
(1985) for the case of atom-molecule collisions. As discussed by these
authors, the $\Delta J=\Delta F$ propensity is independent of dynamics
and follows properties of Wigner-$6j$ symbols (see Section~2.3). It is
thus expected to become increasingly strong as $J$ and $J'$ increase,
as indeed observed in Tables~5$-$8. The fact that it is stronger for
electrons than for He atoms is simply due to the smallness of
electron-impact rates for transitions with $\Delta J>2$.

It should be noted that in the case of He$-$N$_2$H$^+$ collisions,
Daniel et al. (2005) have found that the $\Delta J=\Delta F$
propensity does not always apply at low energy, owing to the presence
of closed-channel resonances. The importance of such effects for
electron-molecule collisions is however unclear and would deserve
further studies.

\begin{table*}
\begin{center}
\caption{Hyperfine de-excitation rates in cm$^3$s$^{-1}$ for
  HCN. Where necessary powers of ten are given in parentheses.}
\begin{tabular}{lllllll}

\hline
J & F & J' & F' & 10~K & 100~K & 1000~K \\
\hline
1 & 0 & 0 & 1 & 0.373(-5) & 0.297(-5) & 0.173(-5) \\
1 & 1 & 0 & 1 & 0.373(-5) & 0.297(-5) & 0.173(-5) \\
1 & 1 & 1 & 0 & 0.0 & 0.0 & 0.0 \\
1 & 2 & 0 & 1 & 0.373(-5) & 0.297(-5) & 0.173(-5) \\
1 & 2 & 1 & 0 & 0.492(-7) & 0.277(-7) & 0.121(-7) \\
1 & 2 & 1 & 1 & 0.111(-6) & 0.624(-7) & 0.272(-7) \\
2 & 1 & 0 & 1 & 0.123(-6) & 0.693(-7) & 0.302(-7) \\
2 & 1 & 1 & 0 & 0.199(-5) & 0.164(-5) & 0.101(-5) \\
2 & 1 & 1 & 1 & 0.149(-5) & 0.123(-5) & 0.757(-6) \\
2 & 1 & 1 & 2 & 0.110(-6) & 0.887(-7) & 0.533(-7) \\
2 & 2 & 0 & 1 & 0.123(-6) & 0.693(-7) & 0.302(-7) \\
2 & 2 & 1 & 0 & 0.0 & 0.0 & 0.0 \\
2 & 2 & 1 & 1 & 0.269(-5) & 0.221(-5) & 0.136(-5) \\
2 & 2 & 1 & 2 & 0.902(-6) & 0.740(-6) & 0.456(-6) \\
2 & 2 & 2 & 1 & 0.615(-7) & 0.346(-7) & 0.151(-7) \\
2 & 3 & 0 & 1 & 0.123(-6) & 0.693(-7) & 0.302(-7) \\
2 & 3 & 1 & 0 & 0.259(-8) & 0.165(-8) & 0.683(-9) \\
2 & 3 & 1 & 1 & 0.518(-8) & 0.331(-8) & 0.137(-8) \\
2 & 3 & 1 & 2 & 0.358(-5) & 0.295(-5) & 0.182(-5) \\
2 & 3 & 2 & 1 & 0.524(-8) & 0.297(-8) & 0.128(-8) \\
2 & 3 & 2 & 2 & 0.504(-7) & 0.284(-7) & 0.124(-7) \\
3 & 2 & 0 & 1 & 0.756(-8) & 0.467(-8) & 0.182(-8) \\
3 & 2 & 1 & 0 & 0.652(-7) & 0.398(-7) & 0.179(-7) \\
3 & 2 & 1 & 1 & 0.652(-7) & 0.398(-7) & 0.179(-7) \\
3 & 2 & 1 & 2 & 0.961(-8) & 0.591(-8) & 0.264(-8) \\
3 & 2 & 2 & 1 & 0.277(-5) & 0.232(-5) & 0.150(-5) \\
3 & 2 & 2 & 2 & 0.516(-6) & 0.433(-6) & 0.279(-6) \\
3 & 2 & 2 & 3 & 0.157(-7) & 0.130(-7) & 0.826(-8) \\
3 & 3 & 0 & 1 & 0.756(-8) & 0.467(-8) & 0.182(-8) \\
3 & 3 & 1 & 0 & 0.0 & 0.0 & 0.0 \\
3 & 3 & 1 & 1 & 0.932(-7) & 0.570(-7) & 0.256(-7) \\
3 & 3 & 1 & 2 & 0.468(-7) & 0.286(-7) & 0.129(-7) \\
3 & 3 & 2 & 1 & 0.260(-8) & 0.169(-8) & 0.732(-9) \\
3 & 3 & 2 & 2 & 0.293(-5) & 0.246(-5) & 0.159(-5) \\
3 & 3 & 2 & 3 & 0.369(-6) & 0.310(-6) & 0.200(-6) \\
3 & 3 & 3 & 2 & 0.353(-7) & 0.199(-7) & 0.866(-8) \\
3 & 4 & 0 & 1 & 0.756(-8) & 0.467(-8) & 0.182(-8) \\
3 & 4 & 1 & 0 & 0.777(-10) & 0.554(-10) & 0.207(-10) \\
3 & 4 & 1 & 1 & 0.146(-9) & 0.104(-9) & 0.388(-10) \\
3 & 4 & 1 & 2 & 0.140(-6) & 0.854(-7) & 0.384(-7) \\
3 & 4 & 2 & 1 & 0.288(-9) & 0.188(-9) & 0.813(-10) \\
3 & 4 & 2 & 2 & 0.240(-8) & 0.157(-8) & 0.677(-9) \\
3 & 4 & 2 & 3 & 0.330(-5) & 0.277(-5) & 0.179(-5) \\
3 & 4 & 3 & 2 & 0.134(-8) & 0.759(-9) & 0.329(-9) \\
3 & 4 & 3 & 3 & 0.286(-7) & 0.161(-7) & 0.702(-8) \\
\hline
\end{tabular}
\end{center}
\end{table*}

\begin{table*}
\begin{center}
\caption{Hyperfine de-excitation rates in cm$^3$s$^{-1}$ for DCN. Where
  necessary powers of ten are given in parentheses.}
\begin{tabular}{lllllll}

\hline
J & F & J' & F' & 10~K & 100~K & 1000~K \\
\hline
1 & 0 & 0 & 1 & 0.392(-5) & 0.311(-5) & 0.179(-5) \\
1 & 1 & 0 & 1 & 0.392(-5) & 0.311(-5) & 0.179(-5) \\
1 & 1 & 1 & 0 & 0.0 & 0.0 & 0.0 \\
1 & 2 & 0 & 1 & 0.392(-5) & 0.311(-5) & 0.179(-5) \\
1 & 2 & 1 & 0 & 0.516(-7) & 0.281(-7) & 0.121(-7) \\
1 & 2 & 1 & 1 & 0.116(-6) & 0.632(-7) & 0.273(-7) \\
2 & 1 & 0 & 1 & 0.129(-6) & 0.702(-7) & 0.303(-7) \\
2 & 1 & 1 & 0 & 0.212(-5) & 0.172(-5) & 0.105(-5) \\
2 & 1 & 1 & 1 & 0.159(-5) & 0.129(-5) & 0.786(-6) \\
2 & 1 & 1 & 2 & 0.117(-6) & 0.933(-7) & 0.553(-7) \\
2 & 2 & 0 & 1 & 0.129(-6) & 0.702(-7) & 0.303(-7) \\
2 & 2 & 1 & 0 & 0.0 & 0.0 & 0.0 \\
2 & 2 & 1 & 1 & 0.286(-5) & 0.233(-5) & 0.142(-5) \\
2 & 2 & 1 & 2 & 0.960(-6) & 0.780(-6) & 0.474(-6) \\
2 & 2 & 2 & 1 & 0.645(-7) & 0.351(-7) & 0.152(-7) \\
2 & 3 & 0 & 1 & 0.129(-6) & 0.702(-7) & 0.303(-7) \\
2 & 3 & 1 & 0 & 0.277(-8) & 0.170(-8) & 0.689(-9) \\
2 & 3 & 1 & 1 & 0.554(-8) & 0.341(-8) & 0.138(-8) \\
2 & 3 & 1 & 2 & 0.381(-5) & 0.310(-5) & 0.189(-5) \\
2 & 3 & 2 & 1 & 0.550(-8) & 0.301(-8) & 0.129(-8) \\
2 & 3 & 2 & 2 & 0.529(-7) & 0.288(-7) & 0.124(-7) \\
3 & 2 & 0 & 1 & 0.798(-8) & 0.478(-8) & 0.183(-8) \\
3 & 2 & 1 & 0 & 0.685(-7) & 0.406(-7) & 0.180(-7) \\
3 & 2 & 1 & 1 & 0.685(-7) & 0.406(-7) & 0.180(-7) \\
3 & 2 & 1 & 2 & 0.101(-7) & 0.602(-8) & 0.266(-8) \\
3 & 2 & 2 & 1 & 0.298(-5) & 0.247(-5) & 0.156(-5) \\
3 & 2 & 2 & 2 & 0.556(-6) & 0.459(-6) & 0.290(-6) \\
3 & 2 & 2 & 3 & 0.169(-7) & 0.138(-7) & 0.857(-8) \\
3 & 3 & 0 & 1 & 0.798(-8) & 0.478(-8) & 0.183(-8) \\
3 & 3 & 1 & 0 & 0.0 & 0.0 & 0.0 \\
3 & 3 & 1 & 1 & 0.979(-7) & 0.581(-7) & 0.258(-7) \\
3 & 3 & 1 & 2 & 0.491(-7) & 0.291(-7) & 0.129(-7) \\
3 & 3 & 2 & 1 & 0.280(-8) & 0.175(-8) & 0.739(-9) \\
3 & 3 & 2 & 2 & 0.315(-5) & 0.261(-5) & 0.165(-5) \\
3 & 3 & 2 & 3 & 0.397(-6) & 0.328(-6) & 0.207(-6) \\
3 & 3 & 3 & 2 & 0.370(-7) & 0.202(-7) & 0.869(-8) \\
3 & 4 & 0 & 1 & 0.798(-8) & 0.478(-8) & 0.183(-8) \\
3 & 4 & 1 & 0 & 0.825(-10) & 0.574(-10) & 0.209(-10) \\
3 & 4 & 1 & 1 & 0.155(-9) & 0.108(-9) & 0.392(-10) \\
3 & 4 & 1 & 2 & 0.147(-6) & 0.870(-7) & 0.386(-7) \\
3 & 4 & 2 & 1 & 0.312(-9) & 0.195(-9) & 0.821(-10) \\
3 & 4 & 2 & 2 & 0.260(-8) & 0.162(-8) & 0.684(-9) \\
3 & 4 & 2 & 3 & 0.355(-5) & 0.294(-5) & 0.186(-5) \\
3 & 4 & 3 & 2 & 0.141(-8) & 0.769(-9) & 0.330(-9) \\
3 & 4 & 3 & 3 & 0.300(-7) & 0.163(-7) & 0.705(-8) \\
\hline
\end{tabular}
\end{center}
\end{table*}


\begin{table*}
\begin{center}
\caption{Hyperfine de-excitation rates cm$^3$s$^{-1}$ for HNC. Where
  necessary powers of ten are given in parentheses.}
\begin{tabular}{lllllll}

\hline
J & F & J' & F' & 10~K & 100~K & 1000~K \\
\hline
1 & 0 & 0 & 1 & 0.388(-5) & 0.307(-5) & 0.177(-5) \\
1 & 1 & 0 & 1 & 0.388(-5) & 0.307(-5) & 0.177(-5) \\
1 & 1 & 1 & 0 & 0.0 & 0.0 & 0.0 \\
1 & 2 & 0 & 1 & 0.388(-5) & 0.307(-5) & 0.177(-5) \\
1 & 2 & 1 & 0 & 0.347(-7) & 0.203(-7) & 0.115(-7) \\
1 & 2 & 1 & 1 & 0.780(-7) & 0.456(-7) & 0.259(-7) \\
2 & 1 & 0 & 1 & 0.867(-7) & 0.507(-7) & 0.288(-7) \\
2 & 1 & 1 & 0 & 0.205(-5) & 0.167(-5) & 0.103(-5) \\
2 & 1 & 1 & 1 & 0.154(-5) & 0.126(-5) & 0.770(-6) \\
2 & 1 & 1 & 2 & 0.111(-6) & 0.893(-7) & 0.543(-7) \\
2 & 2 & 0 & 1 & 0.867(-7) & 0.507(-7) & 0.288(-7) \\
2 & 2 & 1 & 0 & 0.0 & 0.0 & 0.0 \\
2 & 2 & 1 & 1 & 0.277(-5) & 0.226(-5) & 0.139(-5) \\
2 & 2 & 1 & 2 & 0.928(-6) & 0.757(-6) & 0.464(-6) \\
2 & 2 & 2 & 1 & 0.434(-7) & 0.254(-7) & 0.144(-7) \\
2 & 3 & 0 & 1 & 0.867(-7) & 0.507(-7) & 0.288(-7) \\
2 & 3 & 1 & 0 & 0.199(-8) & 0.134(-8) & 0.719(-9) \\
2 & 3 & 1 & 1 & 0.398(-8) & 0.267(-8) & 0.144(-8) \\
2 & 3 & 1 & 2 & 0.369(-5) & 0.302(-5) & 0.185(-5) \\
2 & 3 & 2 & 1 & 0.373(-8) & 0.220(-8) & 0.123(-8) \\
2 & 3 & 2 & 2 & 0.356(-7) & 0.208(-7) & 0.118(-7) \\
3 & 2 & 0 & 1 & 0.586(-8) & 0.381(-8) & 0.193(-8) \\
3 & 2 & 1 & 0 & 0.459(-7) & 0.293(-7) & 0.172(-7) \\
3 & 2 & 1 & 1 & 0.459(-7) & 0.293(-7) & 0.172(-7) \\
3 & 2 & 1 & 2 & 0.683(-8) & 0.438(-8) & 0.254(-8) \\
3 & 2 & 2 & 1 & 0.284(-5) & 0.237(-5) & 0.152(-5) \\
3 & 2 & 2 & 2 & 0.528(-6) & 0.440(-6) & 0.282(-6) \\
3 & 2 & 2 & 3 & 0.158(-7) & 0.131(-7) & 0.836(-8) \\
3 & 3 & 0 & 1 & 0.586(-8) & 0.381(-8) & 0.193(-8) \\
3 & 3 & 1 & 0 & 0.0 & 0.0 & 0.0 \\
3 & 3 & 1 & 1 & 0.657(-7) & 0.419(-7) & 0.246(-7) \\
3 & 3 & 1 & 2 & 0.330(-7) & 0.210(-7) & 0.123(-7) \\
3 & 3 & 2 & 1 & 0.198(-8) & 0.136(-8) & 0.767(-9) \\
3 & 3 & 2 & 2 & 0.300(-5) & 0.250(-5) & 0.161(-5) \\
3 & 3 & 2 & 3 & 0.378(-6) & 0.315(-6) & 0.202(-6) \\
3 & 3 & 3 & 2 & 0.249(-7) & 0.146(-7) & 0.827(-8) \\
3 & 4 & 0 & 1 & 0.586(-8) & 0.381(-8) & 0.193(-8) \\
3 & 4 & 1 & 0 & 0.699(-10) & 0.522(-10) & 0.229(-10) \\
3 & 4 & 1 & 1 & 0.131(-9) & 0.978(-10) & 0.429(-10) \\
3 & 4 & 1 & 2 & 0.985(-7) & 0.628(-7) & 0.368(-7) \\
3 & 4 & 2 & 1 & 0.220(-9) & 0.151(-9) & 0.852(-10) \\
3 & 4 & 2 & 2 & 0.184(-8) & 0.126(-8) & 0.710(-9) \\
3 & 4 & 2 & 3 & 0.338(-5) & 0.282(-5) & 0.181(-5) \\
3 & 4 & 3 & 2 & 0.952(-9) & 0.560(-9) & 0.315(-9) \\
3 & 4 & 3 & 3 & 0.202(-7) & 0.118(-7) & 0.670(-8) \\
\hline
\end{tabular}
\end{center}
\end{table*}

\begin{table*}
\begin{center}
\caption{Hyperfine de-excitation rates cm$^3$s$^{-1}$ for DNC. Where
  necessary powers of ten are given in parentheses.}
\begin{tabular}{lllllll}

\hline
J & F & J' & F' & 10~K & 100~K & 1000~K \\
\hline
1 & 0 & 0 & 1 & 0.408(-5) & 0.321(-5) & 0.183(-5) \\
1 & 1 & 0 & 1 & 0.408(-5) & 0.321(-5) & 0.183(-5) \\
1 & 1 & 1 & 0 & 0.0 & 0.0 & 0.0 \\
1 & 2 & 0 & 1 & 0.408(-5) & 0.321(-5) & 0.183(-5) \\
1 & 2 & 1 & 0 & 0.360(-7) & 0.205(-7) & 0.115(-7) \\
1 & 2 & 1 & 1 & 0.811(-7) & 0.462(-7) & 0.259(-7) \\
2 & 1 & 0 & 1 & 0.901(-7) & 0.513(-7) & 0.288(-7) \\
2 & 1 & 1 & 0 & 0.218(-5) & 0.176(-5) & 0.106(-5) \\
2 & 1 & 1 & 1 & 0.163(-5) & 0.132(-5) & 0.799(-6) \\
2 & 1 & 1 & 2 & 0.118(-6) & 0.939(-7) & 0.563(-7) \\
2 & 2 & 0 & 1 & 0.901(-7) & 0.513(-7) & 0.288(-7) \\
2 & 2 & 1 & 0 & 0.0 & 0.0 & 0.0 \\
2 & 2 & 1 & 1 & 0.294(-5) & 0.238(-5) & 0.144(-5) \\
2 & 2 & 1 & 2 & 0.986(-6) & 0.797(-6) & 0.481(-6) \\
2 & 2 & 2 & 1 & 0.450(-7) & 0.256(-7) & 0.144(-7) \\
2 & 3 & 0 & 1 & 0.901(-7) & 0.513(-7) & 0.288(-7) \\
2 & 3 & 1 & 0 & 0.211(-8) & 0.137(-8) & 0.726(-9) \\
2 & 3 & 1 & 1 & 0.421(-8) & 0.274(-8) & 0.145(-8) \\
2 & 3 & 1 & 2 & 0.392(-5) & 0.318(-5) & 0.192(-5) \\
2 & 3 & 2 & 1 & 0.388(-8) & 0.223(-8) & 0.123(-8) \\
2 & 3 & 2 & 2 & 0.370(-7) & 0.211(-7) & 0.118(-7) \\
3 & 2 & 0 & 1 & 0.614(-8) & 0.388(-8) & 0.194(-8) \\
3 & 2 & 1 & 0 & 0.479(-7) & 0.297(-7) & 0.172(-7) \\
3 & 2 & 1 & 1 & 0.479(-7) & 0.297(-7) & 0.172(-7) \\
3 & 2 & 1 & 2 & 0.713(-8) & 0.445(-8) & 0.254(-8) \\
3 & 2 & 2 & 1 & 0.304(-5) & 0.251(-5) & 0.158(-5) \\
3 & 2 & 2 & 2 & 0.566(-6) & 0.467(-6) & 0.293(-6) \\
3 & 2 & 2 & 3 & 0.170(-7) & 0.139(-7) & 0.868(-8) \\
3 & 3 & 0 & 1 & 0.614(-8) & 0.388(-8) & 0.194(-8) \\
3 & 3 & 1 & 0 & 0.0 & 0.0 & 0.0 \\
3 & 3 & 1 & 1 & 0.686(-7) & 0.425(-7) & 0.246(-7) \\
3 & 3 & 1 & 2 & 0.344(-7) & 0.213(-7) & 0.123(-7) \\
3 & 3 & 2 & 1 & 0.212(-8) & 0.140(-8) & 0.774(-9) \\
3 & 3 & 2 & 2 & 0.321(-5) & 0.265(-5) & 0.167(-5) \\
3 & 3 & 2 & 3 & 0.404(-6) & 0.334(-6) & 0.210(-6) \\
3 & 3 & 3 & 2 & 0.259(-7) & 0.147(-7) & 0.827(-8) \\
3 & 4 & 0 & 1 & 0.614(-8) & 0.388(-8) & 0.194(-8) \\
3 & 4 & 1 & 0 & 0.737(-10) & 0.536(-10) & 0.231(-10) \\
3 & 4 & 1 & 1 & 0.138(-9) & 0.100(-9) & 0.432(-10) \\
3 & 4 & 1 & 2 & 0.103(-6) & 0.636(-7) & 0.368(-7) \\
3 & 4 & 2 & 1 & 0.235(-9) & 0.156(-9) & 0.860(-10) \\
3 & 4 & 2 & 2 & 0.196(-8) & 0.130(-8) & 0.717(-9) \\
3 & 4 & 2 & 3 & 0.362(-5) & 0.299(-5) & 0.188(-5) \\
3 & 4 & 3 & 2 & 0.990(-9) & 0.567(-9) & 0.315(-9) \\
3 & 4 & 3 & 3 & 0.210(-7) & 0.120(-7) & 0.670(-8) \\
\hline
\end{tabular}
\end{center}
\end{table*}

\begin{table*}
\begin{center}
\caption{Relative propensities for hyperfine collisional rates as a
  percentage of the total rates for $T=20$~K. Present=rates calculated
  here; He (CC)=close-coupling He$-$HCN results of Monteiro \& Stutzki
  (1986); He (IOS)=IOS scaling for He$-$HCN.}
\begin{tabular}{llllrrr}
\hline
J & F & J' & F' & Present & He (CC) & He (IOS) \\
\hline
1 & 1 & 2 & 1 & 24.9 & 21.4 & 19.2 \\
  &   & 2 & 2 & 74.9 & 59.7 & 65.4 \\
  &   & 2 & 3 & 0.2 & 18.9 & 15.4 \\
1 & 1 & 3 & 2 & 33.3 & 30.0 & 26.3 \\
  &   & 3 & 3 & 66.6 & 61.3 & 60.5 \\
  &   & 3 & 4 & 0.1 & 8.8 & 13.2 \\
2 & 2 & 3 & 2 & 11.2 & 17.0 & 16.6 \\
  &   & 3 & 3 & 88.7 & 69.8 & 75.8 \\
  &   & 3 & 4 & 0.1 & 13.3 & 7.7 \\
\hline
\end{tabular}
\end{center}
\end{table*}

It can be noticed in Tables~5$-$8 that in addition to the $\Delta
J=\Delta F$ propensity rule, there is a rigorous selection rule at the
IOS level: the $(J, F=J) \rightarrow (J'=1, F'=0)$ transitions are
strictly forbidden, following properties of Wigner-$6j$
symbols. Monteiro \& Stutzki (1986) have shown that for He$-$HCN this
selection rule becomes only a propensity rule at the close-coupling
level.
\newline
Finally we note that the influence of collisional processes on HCN
hyperfine anomalies is still a matter of debate in the literature
(e.g. Gonz\'ales-Alfonso \& Cernicharo 1993; Turner 2001; Daniel,
Cernicharo \& Dubernet 2006). Future radiative transfer studies
including the present hyperfine rates will thus help to further assess
the role of collision-induced transitions between HCN hyperfine
levels. In this context, it is important to remember that H$_2$$-$HCN
collisional rates are currently obtained by scaling He$-$HCN rates by
the reduced mass ratio. While for para-H$_2$($J$=0) this assumption is
probably qualitatively correct, rates for H$_2(J\neq 0)$ might be
significantly different, as found for example in the case of
H$_2-$H$_2$O (Faure et al. 2007 and references therein). Furthermore,
it is worth stressing that rates for excitation of HCN by H atoms and
H$_2$O molecules are completely unknown.

\section{Conclusions}

We have calculated electron-impact rotational and hyperfine
(de)excitation rates for isotopologues of HCN and HNC. These
calculations are based on the molecular {\bf R}-matrix method combined
with the (threshold corrected) ANR approximation. Our results show
that such collisions are essentially dominated by dipolar transitions
owing to the large dipole moments of the considered species. However,
short-range effects are important and were included via {\bf R}-matrix
wavefunctions as corrections to the Born approximation. Dipole
forbidden transitions have thus appreciable rates which cannot be
neglected in any detailed population models of isotopologues of HCN
and HNC. In particular, we have shown that electron-impact rates are
crucial for modelling environments where the electron fraction is
larger than 10$^{-6}$. In this context, we note the recent suggestion
by Jimenez-Serra et al. (2006) to use rotational emissions of HCO$^+$,
HCN and HNC to probe electron densities in C-type shocks. The present
rates should help to investigate in more detail the electron density
enhancements expected during the first stages of a C-type shock
evolution. Finally, our work might help to understand and to model the
observed variable HCN/HNC ratio in comets (Biver et al. 2006 and
references therein).

\label{lastpage}

\end{document}